\documentclass[11pt]{article}
\usepackage{amsmath,amssymb,color,epsfig,cite,CJK}
\usepackage{graphicx}
\usepackage{subfigure}
\usepackage{setspace}
\usepackage{braket}

\usepackage{amsfonts}

\newcommand{\hoch}[1]{$\, ^{#1}$}

\makeatletter
\@addtoreset{equation}{section}
\makeatother

\newcommand{\be}{\begin{equation}}
\newcommand{\ee}{\end{equation}}
\newcommand{\bea}{\setlength\arraycolsep{2pt} \begin{eqnarray}}
\newcommand{\eea}{\end{eqnarray}}
\newcommand{\nn}{\nonumber}

\newcommand{\bpm}{\begin{pmatrix}}
\newcommand{\epm}{\end{pmatrix}}

\def\ft#1#2{{\textstyle{\frac{\scriptstyle #1}{\scriptstyle #2} } }}
\def\fft#1#2{{\frac{#1}{#2}}}

\def\0{{\sst{(0)}}}
\def\1{{\sst{(1)}}}
\def\2{{\sst{(2)}}}
\def\3{{\sst{(3)}}}
\def\4{{\sst{(4)}}}
\def\5{{\sst{(5)}}}
\def\6{{\sst{(6)}}}
\def\7{{\sst{(7)}}}
\def\8{{\sst{(8)}}}
\def\sst#1{{\scriptscriptstyle #1}}

\begin{document}


\begin{center}
{\large {\bf On the Size of a Black Hole: The Schwarzschild is the Biggest}}

\vspace{10pt}
H. L\"u\hoch{1} and Hong-Da Lyu\hoch{2}

\vspace{10pt}

\hoch{1}{\it Center for Joint Quantum Studies and Department of Physics,\\
School of Science, Tianjin University, Tianjin 300350, China}

\vspace{10pt}

\hoch{2}{\it Physics Department, Beijing Normal University, Beijing 100875, China}

\vspace{40pt}

\underline{ABSTRACT}
\end{center}

We consider static black holes in Einstein gravity and study parameters characterizing the black hole size, namely the radii of the horizon $R_+$, photon sphere $R_{\rm ph}$ and black hole shadow $R_{\rm sh}$.  We find a sequence of inequalities $\fft32 R_+\, \le\, R_{\rm ph}\, \le \,\fft{1}{\sqrt3} R_{\rm sh}\,\le\, 3 M$, where $M$ is the black hole mass. These are consistent with and beyond the previously known upper bounds in literature.  The Schwarzschild black hole saturates all the inequalities, making it the biggest of all for given mass. The inequalities include an upper bound of entropy for any quantum system with given energy. We also point out that some black holes satisfying the dominant energy condition can trap photons to form a stable photon shield outside the event horizon, but the shadow hides it from an observer at infinity.

\vfill {\footnotesize  mrhonglu@gmail.com, \ \ \ hongda\_lv@163.com}

\thispagestyle{empty}

\pagebreak

\tableofcontents
\addtocontents{toc}{\protect\setcounter{tocdepth}{2}}

\newpage

\section{Introduction}

Black holes are fundamental objects in Einstein's gravity and they are completely specified by a few conserved quantities such as the mass, charge and angular momentum.  The interior and outside of a black hole are separated by the event horizon, whose area provides a measure of the geometric size of a black hole. However, at least classically, it is not feasible to detect the black hole horizon directly, either locally or from asymptotic infinity. For an observer at the asymptotic infinity, properties of black holes are extracted from analysing the motions of the surrounding objects, which in many circumstances can be viewed as test particles in geodesic motions.

Massless photons are perhaps the most convenient test particles for the purpose of observation.  It can be easily demonstrated that they can form a photon sphere outside a black hole horizon.  The photon sphere is typically unstable and can cast a ``shadow'' for an observer at the asymptotic infinity. Since the earlier works on the Schwarzschild shadows \cite{Synge:1966okc,Luminet:1979nyg}, the subject on black hole photon spheres and shadows has been intensively investigated [3-35]. Recently, a picture of such a black hole shadow was taken \cite{Akiyama:2019cqa} for the first time, giving a direct impression of the appearance of the black hole size and shape.

With the advancement of the astrophysical observations, it is of great interest to study the relations among parameters measuring the size of a black hole and how they are determined by the black hole mass.  These parameters include those associated with the black hole horizons, photon spheres and shadows.
In this paper, we consider only the spherically symmetric and static black holes in four dimensions, in which case the horizons, photon spheres are all round spheres.  They can be specified by their radii $R_+$ and $R_{\rm ph}$ respectively.  The black hole shadow is a disk specified by its radius $R_{\rm sh}$.  If the horizon radius gives the ``actual'' size of a black hole, $R_{\rm ph}$ and $R_{\rm sh}$ are measures of its ``apparent'' size.

Hod proved that for Einstein gravity coupled to matter satisfying the dominant energy condition, together with negative trace of the energy-momentum tensor, the radius of the (unstable) photon sphere outside the black hole of mass $M$ has an upper bound, namely \cite{Hod:2017xkz}
\be
R_{\rm ph} \le 3M\,.\label{known1}
\ee
The same energy conditions also imply that \cite{Cvetic:2016bxi}
\be
R_{\rm ph}\le \fft1{\sqrt3}\,R_{\rm sh}\,.\label{known2}
\ee
We generalize these theorems and propose a sequence of inequalities for the aforementioned size parameters of a black hole:
\be
\ft32 R_+\, \le\, R_{\rm ph}\, \le \,\fft{R_{\rm sh}}{\sqrt3} \,\le\, 3 M.\label{conjecture}
\ee
Hod also made a conjecture of a lower bound for $R_{\rm sh}$, namely $R_{\rm sh}\ge 2M$ \cite{Hod:2012nk}, but counterexample was found in \cite{Cvetic:2016bxi}. In (\ref{conjecture}), we propose instead an upper bound for the shadow radius.

The purpose of this paper is to study a diverse selection of black holes and verify the above relations. We include scalar hairy black holes that satisfy the null energy condition, but do not satisfy the dominant or even the weak energy conditions. These tests thus go beyond the conditions that were used to prove (\ref{known1}) and (\ref{known2}). The paper is organized as follows.  In section \ref{sec:formalism}, we present the general formalism for calculating the aforementioned radii in terms of the metric functions. In sections 3,4,5,6,7 we verify our conjecture (\ref{conjecture}) for a variety of black holes in Einstein gravity.  We conclude and discuss in section 8.

\section{General formalism and statements}
\label{sec:formalism}

We begin with setting up the general formalism and statements.  We consider general static and spherically symmetric black holes that are asymptotic to Minkowski spacetime.  The general ansatz of the metric takes the form
\be
ds^2 = - h(r) dt^2 + \fft{dr^2}{f(r)} + R(r)^2 (d\theta^2 + \sin^2\theta d\phi^2)\,.
\ee
Here $R(r)$ is the radius of the foliating sphere. We would like to point out that we can always choose appropriate radial coordinate gauge and set $R(r)=r$ or alternatively $h=f$. Since we shall study a variety of explicit black hole examples later, we shall leave $(h,f,R)$ as three generic functions to cope with the different coordinate choices.  However, we do choose the coordinate gauge such that the asymptotic infinity $R\rightarrow \infty$ arises as $r\rightarrow +\infty$.  For asymptotically flat spacetimes, we must have
\be
h=1 - \fft{2M}{R} + {\cal O}\big(\fft{1}{R^2}\big)\,,
\ee
where $M$ is the mass. For the solution to describe a black hole, there exists an event horizon $r_+$ such that $f(r_+)=h(r_+)=0$ and $r_+$ is the largest root and it can either be a single or double root.  The latter gives rise to an extremal black hole.  The horizon radius is then given by
\be
R_+=R(r_+)\,.
\ee

There typically exists an (unstable) photon sphere outside a black hole and it will give rise to a black hole shadow for an observer at infinity. We shall now give the formulae of the radii of both the photon sphere and black hole shadow, whose shape is a disk for a spherically symmetric black hole. For a massless particle like photon, its geometric motion is invariant under any conformal transformation of the background metric.  It is thus equivalent and advantageous to consider
\be
ds^2 = - dt^2 + dr_*^2 + \tilde R^2 (d\theta^2 + \sin^2\theta d\phi^2)\,,
\ee
where
\be
dr_* = \fft{dr}{\sqrt{hf}}\,,\qquad \tilde R = \fft{R}{\sqrt{h}}\,.
\ee
Note that $\tilde R$ was referred to as the optical radius $R_{\rm opt}$ in \cite{Cvetic:2016bxi}. The Lagrangian of the geodesic motion of a photon is then given by
\be
L=\ft12 \big(-\dot t^2 + \dot r_*^2 + \tilde R^2 (\dot \theta^2 + \sin^2\theta\, \dot \phi^2)\big)\,,
\ee
subject to the null constraint $L=0$. Here a dot denotes a derivative with respect to the affine parameter.
Thus we have two ignorable coordinates $(t, \phi)$ giving rise to two conserved quantities $(E, \ell_\phi)$, the energy and angular momentum. To be specific, we have
\be
\dot t=E\,,\qquad \dot \phi = \fft{\ell_\phi}{\tilde R^2\sin^2\theta}\,.
\ee
The $\theta$ equation of motion
\be
\tilde R^2 \Big(\tilde R^2 \dot \theta \Big)\dot{} = \fft{\ell_\phi^2\, \cos\theta}{\sin^3\theta}\,,
\ee
can be integrated to
\be
\dot \theta = \fft{\sqrt{\kappa - \ell_\phi^2 \cot^2\theta}}{\tilde R^2 }\,.
\ee
Here $\kappa$ is the integration constant of the first integral.  The first integral of the radial coordinate becomes
\be
H=\ft12 \dot r_*^2 + V_{\rm eff} = \ft12 E^2\,,\qquad V_{\rm eff} = \fft{\kappa + \ell_\phi^2}{2 \tilde R^2}\,.
\ee
Thus there is a photon sphere for which $r_*=$constant at certain $r=r_{\rm ph}$ if we have $V_{\rm eff}'=0$. In other words, the photon sphere and its radius $R_{\rm ph}$ are determined by
\be
\big(\fft{h}{R^2}\big)'\Big|_{r_{\rm ph}} =0\,,\qquad R_{\rm ph} = R(r_{\rm ph})\,.
\ee
Note that for the photon sphere, have
\be
\fft{\kappa +\ell_{\phi}^2}{E^2}=\tilde R^2\Big|_{r_{\rm ph}}\,.\label{psenergy}
\ee

The unstable photon sphere will cast a shadow on the observer's sky, which is is plane perpendicular to the line joining the black hole with an asymptotic observer \cite{Vazquez:2003zm}.  For spherically-symmetric black holes, the shadow is a disk of radius $R_{\rm sh}$, given by
\bea
R_{\rm sh} &=& \lim_{r\rightarrow \infty} \tilde R^2 \sqrt{\big(\fft{d\theta}{dr}\big)^2 + \sin^2\theta\, \big(\fft{d\phi}{dr}\big)^2}\nn\\
&=& \sqrt{\fft{\kappa + \ell_{\phi}^2}{E}} = \tilde R\Big|_{r_{\rm ph}}\,.
\eea
In other words we have
\be
R_{\rm sh} = \fft{R_{\rm ph}}{\sqrt{h(r_{\rm ph})}}\,.
\ee

We now consider a well studied example, namely the Schwarzschild black hole, whose metric functions are
\be
h=f= 1- \fft{2M}{r}\,,\qquad R=r\,.
\ee
We have
\be
R_+=2M\,,\qquad R_{\rm ph} = 3M\,,\qquad R_{\rm sh}=3\sqrt3 M\,.
\ee
The Schwarzschild black hole plays an important role in Einstein gravity since it does not involve matter, except at the curvature singularity, where the classical solution breaks down. Its properties can thus be universal sometime or at least provide a bound for general black holes. One such example is the Hod's theorem (\ref{known1}) and the other is (\ref{known2}). These inequalities are related to the metric function inequality $h(r_{\rm rp})\le 1/3$, which can be established with the dominant energy condition, together with negative trace of the energy-momentum tensor \cite{Hod:2017xkz}. In this paper, we propose a sequence of more general inequalities (\ref{conjecture}), and the Schwarzschild black hole saturates them all. We shall not prove the bounds (\ref{conjecture}) abstractly, but validate them with diverse types of asymptotically-flat black holes in four dimensions.  We shall not restrict ourselves with only black holes satisfying the dominant energy condition, but also include those satisfying the more relaxed null energy condition. Thus even for the inequalities (\ref{known1}) and (\ref{known2}), our tests go beyond the original proofs of the theorems. We carry out this task in the remainder of this paper.

\section{Einstein Maxwell theory}
\label{sec:em}

Einstein-Maxwell gravity admits the well-studied Reissner-Nordstr\"om (RN) black hole, with the metric functions
\be
h=f= 1 - \fft{2M}{r} + \fft{Q^2}{r^2}\,,\qquad R=r\,.
\ee
Here $(M,Q)$ are the mass and electric charge of the black hole. Since $r$ is the radius of the foliating spheres, we shall not use the notation $R$ in this section.  The solution describes a black hole when $M\ge Q$, and the black hole is extremal when $M=Q$. Using the formulae in section \ref{sec:formalism}, we can straightforwardly obtain
\bea
r_+&=& M+ \sqrt{M^2-Q^2}\,,\nn\\
r_{\rm ph} &=& \frac{1}{2} \left(\sqrt{9 M^2-8 Q^2}+3 M\right),\nn\\
r_{\rm sh} &=& \frac{\left(\sqrt{9 M^2-8 Q^2}+3 M\right)^2}{2 \sqrt{2 M \left(\sqrt{9 M^2-8 Q^2}+3 M\right)-4 Q^2}}\,.\label{emradii}
\eea
The reality condition implies that $0 \le Q\le M$, with $M=Q$ giving the extremal solution.  It is convenient to parameterize the charge using a dimensionless quantity $x$
\be
Q=\frac{M \sqrt{3 x+1}}{2 x+1}\,,\qquad x\in [0, \infty]\,.\label{emQ}
\ee
Thus $x=0$ gives rise to the extremal black hole while $x=\infty$ reduces the solution to the Schwarzschild one. Substituting the charge expression (\ref{emQ}) into (\ref{emradii}), we find
\bea
\fft{r_{\rm sh}^2}{r_{\rm ph}^2} - 3 &=& \fft{1}{1 + 4x} \ge 0\,,\nn\\
27 - \fft{r_{\rm sh}^2}{M^2} &=& \fft{11 + 72 x + 108x^2}{(1+2x)^2 (1 + 4x)} \ge 0\,,\nn\\
2\ge \fft{r_{\rm ph}}{\rm r_+} &=& \fft32 + \frac{1}{2(8 x+1 + 4 \sqrt{x (4 x+1)})}\ge \fft32\,.
\eea
Thus the RN black hole confirms our conjecture (\ref{conjecture}) with the saturation occurring at $x=\infty$, corresponding to the Schwarzschild solution. Since electric and magnetic charges enter the black hole metric symmetrically, we can draw the same conclusion for the dyonic RN black holes.
\section{Scalar hairy black holes}

In this section, we consider Einstein-scalar gravity with the Lagrangian
\be
{\cal L}=\sqrt{-g} \big({\cal R} - \ft12 (\partial\phi)^2 - V(\phi)\big)\,.
\ee
Note that in order not to be confused with the radius $R$, we use the calligraphic ${\cal R}$ to denote the Ricci scalar. There should also be no confusion between the scalar $\phi$ and the azimuthal angle discussed in section \ref{sec:formalism}. For massless scalar with the concave scalar potential, one can construct asymptotically flat black holes with nontrivial scalar hair.  By scalar hair, we mean that the scalar has the long range asymptotic falloff of order $1/r$.  The reason we would like to study these black holes is that they typically satisfy only the null energy condition, but not the weak or dominant energy conditions. They also have an intriguing feature that the volumes of the black holes can be negative \cite{Liu:2019mxz}. We consider two concrete examples.

\subsection{Example 1}

The scalar potential is given by
\be
V=-2 \alpha ^2 \big(2 \phi -3 \sinh (\phi )+\phi  \cosh (\phi)\big)\,.
\ee
Note that the scalars in this paper are all dimensionless and the coupling constant $\alpha$ has the same dimension as mass. However, the scalar is massless and the leading order of the Taylor expansion of $V$ in small $\phi$ is of order $\phi^5$. This scalar potential is part of the more general scalar potential studied in \cite{Zloshchastiev:2004ny,Zhang:2014sta}. The Einstein-scalar theory admits the spherically-symmetric and static solution, with the metric functions
\bea
h&=&f=1+\alpha ^2 \left(r (q+r) \log \left(\frac{q+r}{r}\right)-\frac{1}{2} q (q+2 r)\right)\,,\nn\\
R &=& \sqrt{r(r+q)}\,,\qquad e^{\phi}=1 + \fft{q}{r}\,.
\eea
The solution contains only one integration constant $q$, parameterizing the mass
\be
M=\ft1{12} \alpha^2 q^3\,.
\ee
For sufficiently large mass, namely
\be
c\equiv \alpha^2 q^2 -2>0\,,\label{esbhcon}
\ee
the solution describes a black hole with one horizon $r_+$, determined by
\be
\fft{2}{\alpha^2} = q \left(q+2 r_+\right)-2 r_+ \left(q+r_+\right) \log \left(\frac{q}{r_+}+1\right).
\ee
The radius of the horizon and the entropy of the black hole are
\be
R_+=\sqrt{r_+(r_+ + q)}\,,\qquad S=\pi R_+^2\,.
\ee
It is straightforward to verify that the first law of the black hole thermodynamics $dM=TdS$ is satisfied, where $T=f'(r_+)/(4\pi)$.

The energy density and pressure outside the horizon is given by
\bea
&&\rho = \fft{1}{4r^2(r+q)^2} \Big(2 \left(q^2+6 q r+6 r^2\right) -\alpha ^2 q^3 (q+2 r) -\left(q^2+12 q r+12 r^2\right)f\Big)\,,\nn\\
&&\rho + p_1= \frac{q^2f}{2 r^2 (q+r)^2}\,,\qquad \rho+p_{2,3}=0\,.
\eea
Thus we see that $\rho$ can be negative and hence it violates the weak energy condition, but the null energy condition is satisfied since we have $f>0$ outside the horizon.  The same conclusion can be made inside the horizon where $-p_1$ becomes the density whilst $(-\rho,p_{2,3})$ are the pressure.

The location of the photon sphere is
\be
r_{\rm ph} = \ft14 q \left(\alpha ^2 q^2-2\right).
\ee
It follows that the radius of the photon sphere and shadow disk are
\be
R_{\rm ph}= \ft14 q\sqrt{(\alpha^2 q^2 +2)(\alpha^2 q^2 -2)}\,,\qquad
R_{\rm sh}= \frac{q}{\sqrt{\alpha ^2 q^2 \log \left(\frac{\alpha ^2 q^2+2}{\alpha ^2 q^2-2}\right)-4}}\,.
\ee
Making use of (\ref{esbhcon}), we find for $c>0$ that
\bea
\fft{M}{R_{\rm sh}}&=& \ft1{12} {(c+2)\sqrt{(c+2)\log (1 + \ft{4}{c}) -4}}\ge \fft{1}{3\sqrt3}\,,\nn\\
\fft{R_{\rm sh}}{R_{\rm ph}} &=& \fft{4}{\sqrt{c(c+4)\Big(
(c+2) \log \left(\frac{4}{c}+1\right)-4
\Big)}}\ge \sqrt3
\eea
The saturation of both inequalities occurs when $c=+\infty$.  To evaluate the range of the ratio $R_{\rm ph}/R_+$, it is advantageous to introduce a positive dimensionless constant $x=r_+/q$.  We find
\be
\fft{R_{\rm ph}}{R_+}=
 \frac{\sqrt{-x (x+1) \log ^2\left(\frac{1}{x}+1\right)+(2 x+1) \log \left(\frac{1}{x}+1\right)-1}}{-2 x^2 \log \left(\frac{1}{x}+1\right)+2 x-2 x \log \left(\frac{1}{x}+1\right)+1}\ge \ft32\,.
\ee
The saturation occurs at $x=+\infty$.  Thus the inequalities in (\ref{conjecture}) are all satisfied.

\subsection{Example 2}

The scalar potential of our second example is
\bea
V&=&\fft{12\alpha}{\mu(4\mu^2-1)}\Bigg(
\sinh \left(\frac{\mu  \phi }{\nu }\right) \left(\left(2 \mu ^2+1\right) \cosh \left(\frac{\phi }{\nu }\right)-2 \mu ^2+2\right)\nn\\
&&\qquad-3 \mu  \sinh \left(\frac{\phi }{\nu }\right) \cosh \left(\frac{\mu  \phi }{\nu }\right)\Bigg),
\eea
where $(\mu,\nu)$ are real constant parameters constrained by $\mu^2 + \nu^2=1$. This is part of the more general potential studied in \cite{Anabalon:2013qua,Feng:2013tza}. The metric functions of the spherically-symmetric and static solutions are \cite{Feng:2013tza}
\bea
h&=&f=H^{-1-\mu} \tilde f\,,\qquad R=r H^{\fft12 (1+\mu)}\,,\qquad H=1+ \fft{q}{r}\,,\nn\\
\tilde f&=&H+ \frac{3 \alpha  H r^2 \left(-H^{2 \mu +1}+H^2 \mu  (2 \mu +1)+H \left(1-4 \mu ^2\right)+\mu  (2 \mu -1)\right)}{\mu  \left(4 \mu ^2-1\right)}\,.
\eea
The metric is asymptotic to the Minkowski spacetime with the mass, parameterized by the integration constant $q$,
\be
M=\ft12 q (\mu + \alpha q^2)\,.
\ee
It should be pointed out that in this particular example the quantity $|\mu|^2\le 1$ is not an integration constant, but a parameter of the theory. The condition for the solution to describe a black hole is
\bea
\mu<\ft12:&&\quad c= \fft{3\alpha q^2}{1-2\mu} -1>0\,,\nn\\
\mu>\ft12:&&\quad \hbox{all}\ \alpha q^2\,.\label{esbhcon2}
\eea
The horizon radius can be determined by
\be
f(r_+)=0\,,\qquad R_{+}=r_+ H^{\fft12 (1+\mu)}(r_+)\,.
\ee
As usual, the scalar hairy black holes can violate the weak energy condition, but they do satisfy the null energy condition. In fact the $\mu=0$ limit yields the example 1 discussed earlier.

The location of the photon sphere can be obtained easily, given by
\be
r_{\rm ph} = \ft12q (3\alpha q^2 +2\mu -1)\,.
\ee
The radii of the photon sphere and the shadow disk are
\bea
R_{\rm ph} &=& \frac{1}{2} q \left(2 \mu +3 \alpha  q^2-1\right)^{\frac{1-\mu }{2}} \left(2 \mu +3 \alpha  q^2+1\right)^{\frac{\mu +1}{2}},\nn\\
R_{\rm sh} &=& \frac{q \sqrt{\mu(4 \mu ^2-1)}}{\sqrt{\left(2 \mu +3 \alpha  q^2-1\right)^{2 \mu } \left(2 \mu +3 \alpha  q^2+1\right)^{-2 \mu } \left(4 \mu +3 \alpha  q^2\right)-3 \alpha  q^2}}\,.
\eea
To analyse the relations among the size parameters $(M, R_{\rm sh}, R_{\rm ph}, R_+)$, we define
\be
X=\fft{27 M^2}{R_{\rm sh}^2}\,,\qquad Y=\fft{R_{\rm sh}^2}{3R_{\rm ph}^2}\,,\qquad
Z=\fft{4R_{\rm ph}^2}{9R_+^2}\,.\label{XYZ}
\ee
Our task is then to prove $X\ge 1$, $Y\ge 1$ and $Z\ge 1$.  We first examine the quantities $(X,Y)$ and they are
\bea
X &=& \frac{27 (\mu +\tilde q)^2 \left((4 \mu +3 \tilde q) \left(1-\frac{2}{2 \mu +3 \tilde q+1}\right)^{2 \mu }-3 \tilde q\right)}{4 \mu  \left(4 \mu ^2-1\right)}\,,\nn\\
Y &=& \frac{4 \mu  \left(4 \mu ^2-1\right) (2 \mu +3 \tilde q-1)^{\mu -1} (2 \mu +3 \tilde q+1)^{-\mu -1}}{3 \left((4 \mu +3 \tilde q) \left(1-\frac{2}{2 \mu +3 \tilde q+1}\right)^{2 \mu }-3 \tilde q\right)}\,,
\eea
where $\tilde q=\alpha q$ is dimensionless.  For sufficiently large $\tilde q$, $(X,Y)$ are monotonically decreasing functions of $\tilde q$, with the large $\tilde q$ expansion
\bea
X &=& 1+\frac{1-\mu ^2}{15 \tilde q^4}-\frac{46 \mu  \left(1-\mu ^2\right)}{405 \tilde q^6}+O\left(\frac{1}{\tilde q^8}\right),\nn\\
Y&=& 1+\frac{2 \left(1-\mu ^2\right)}{45 \tilde q ^4}-\frac{8 \mu  \left(1-\mu ^2\right)}{135 \tilde q ^6}+O\left(\frac{1}{\tilde q^8}\right).
\eea
When $\tilde q$ runs from $+\infty$ to some finite positive value, both $(X,Y)$ increase monotonically.  Three cases emerge depending on the value of $\mu$.  For $\mu >1/2$, $\tilde q$ can approach zero and the maximum value of $(X,Y)$ is
\be
X|_{ \tilde q=0} = \frac{27 \mu ^2}{4 \mu ^2-1}  \left(\frac{2 \mu -1}{2 \mu +1}\right)^{2 \mu }\,,
\qquad
Y|_{\tilde q=0} = \frac{1}{3} \left(\frac{2 \mu +1}{2 \mu -1}\right)^{\mu }\,.
\ee
It is clear that both of these two quantities are greater than 1 for $1/2<\mu <1$.  When $-1\le \mu<\ft12$, the above quantities can be negative and even complex, we should impose the condition (\ref{esbhcon2}).  We find that $(X,Y)$ are decreasing functions of $c$, with
\bea
\left\{X,Y\right\}\Big|_{c\rightarrow 0} &=& \left\{\frac{3 (\mu +1)^2}{4 \mu  (2 \mu +1)},\quad\frac{1}{3c^{1-\mu}} \mu  (2 \mu +1)\right\}\,,\qquad 0<\mu\le \ft12\,,\nn\\
\left\{X,Y\right\}\Big|_{c\rightarrow 0}  &=& \left\{\frac{3 (\mu +1)^2 c^{2 \mu }}{4 \mu  (2 \mu -1)},\quad\mu  (2 \mu -1) c^{-\mu -1}\right\},\qquad
-1\le \mu <0\,.
\eea
Note that the quantities can be divergent when $c\rightarrow 0$, implying that for certain choice of $\mu$, $X$ and $Y$ have no upper bounds.

   To determine the range of $Z$, it is advantageous to define a dimensionless quantity $x=r_+/q$, which runs
from 0 to $+\infty$.  We find
\be
Z=\ft{1}{9} x^2 (x+1)^{-\mu -1} \left(2 \mu +3 \alpha  r_+^2 x^2-1\right)^{1-\mu } \left(2 \mu +3 \alpha  r_+^2 x^2+1\right)^{\mu +1}\,,
\ee
where
\be
3\alpha r_+^2 = \fft{\mu  \left(1-4 \mu ^2\right)}{(2 \mu +1) x (\mu  x+1)+1 - (x+1)^{2 \mu +1}}\,.
\ee
It is easy to establish that $Z$ is a monotonically increasing function of $x$ for $x\in [0, \infty)$, with the minimum $Z=1$ when $x=0$ and the small $x$ of $Z$ is
\be
Z=1+\frac{7}{180} \left(1-\mu ^2\right) x^2-\frac{(1-\mu^2) (63-19 \mu) x^3}{1620}+O\left(x^4\right)\,.
\ee
For large $x$, we have
\be
Z\sim
\left\{
  \begin{array}{ll}
    \frac{1}{9} (2 \mu -1)^{1-\mu } (2 \mu +1)^{\mu +1} x^{1-\mu }, &\qquad  \ft12\le \mu<1 \\
    \frac{1}{9} 2^{\mu +1} (1-2 \mu )^{1-\mu } \mu ^{\mu -1} (2 \mu +1)^{\mu -1} x^{-2 (\mu -1) \mu }, & \qquad
0\le \mu \le \ft12 \\
   \frac{1}{9} 2^{\mu +1} \mu ^{\mu -1} (2 \mu -1)^{1-\mu }, & \qquad -1<\mu\le 0\,.
  \end{array}
\right.
\ee
Thus we establish that $Z\ge 1$.  It follows that the inequalities in (\ref{conjecture}) are all satisfied by all the parameters of the black holes.

\section{Einstein-Maxwell-dilaton theories}

\subsection{Singly-charged black hole}

The Lagrangian of the Einstein-Maxwell-dilaton (EMD) theory is
\be
{\cal L}= \sqrt{-g} ({\cal R} - \ft12 (\partial\phi)^2 - \ft14 e^{a\phi} F^2)\,,\qquad F=dA\,,\qquad a^2=\fft{4}{N} - 1\,.
\ee
The parameter $N$ can take values within $(0,4]$. When $N=1,2,3,4$, the solutions can be embedded in supergravities and lift to become black branes or intersecting branes in string or M-theory
\cite{Duff:1996hp,Cvetic:1996gq}. The metric of the charged black hole solution is
\bea
ds^2 &=& -H^{-\fft12 N} \tilde f dt^2 + H^{\fft12 N} \Big(\fft{dr^2}{\tilde f} + r^2 d\Omega_2^2\Big)\,,\nn\\
\tilde f &=& 1- \fft{\mu}{r}\,,\qquad H=1 + \fft{q}{r}\,.
\eea
In other words, we have
\be
h=\fft{\tilde f}{H^{\fft12N}}\,,\qquad R=r H^{\fft14 N}\,.
\ee
The mass of the solution and the horizon radius are given by
\be
M=\ft12 \mu + \ft14 N q\,,\qquad R_+ = \mu \Big(1 + \fft{q}{\mu}\Big)^{\fft14 N}\,.
\ee
The photon sphere is located at
\be
r_{\rm ph} = \ft14 (3 \mu +(N-2) q) + \ft14 \sqrt{(3 \mu +(N-2) q)^2 -8 \mu (N-3) q}\,.
\ee
Thus the photon sphere and shadow radius are
\bea
R_{\rm ph} &=&
\frac{1}{4} \left(3 \mu +(N-2) q +\sqrt{(3 \mu +(N-2) q)^2-8 \mu  (N-3) q}\right)^{1-\frac{N}{4}} \nn\\
&&\times \left(3 \mu +(N+2) q +\sqrt{(3 \mu +(N-2) q)^2-8 \mu  (N-3) q}\right)^{N/4},\nn\\
R_{\rm sh} &=&
\frac{U}{4 \sqrt{-\mu +(N-2) q +\sqrt{(3 \mu +(N-2) q)^2-8 \mu  (N-3) q}}}\nn\\
U&=& \left(3 \mu +(N-2) q +\sqrt{(3 \mu +(N-2) q)^2-8 \mu  (N-3) q}\right)^{\frac{3-N}{2}} \nn\\
&&\times \left(3 \mu +(N+2) q +\sqrt{(3 \mu +(N-2) q)^2-8 \mu  (N-3) q}\right)^{\fft{N}2}.
\eea
Having obtained the relevant parameters, we are in the position to study the $(X,Y,Z)$ quantities defined in
(\ref{XYZ}).  The $X$ is
\bea
X&=&27 (2 \mu +N q)^2 \left(-\mu +\sqrt{(3 \mu +(N-2) q)^2-8 \mu  (N-3) q}+(N-2) q\right)\nn\\
 &&\times \left(3 \mu +\sqrt{(3 \mu +(N-2) q)^2-8 \mu  (N-3) q}+(N-2) q\right)^{N-3} \nn\\
&&\times \left(3 \mu +\sqrt{(3 \mu +(N-2) q)^2-8 \mu  (N-3) q}+(N+2) q\right)^{-N}.
\eea
For given value of $\mu$, this quantity is a monotonically increasing function of $q$, with the minimum at $q=0$.  The small $q$ expansion is
\be
X=1+\frac{N q}{3 \mu }+\frac{(24-17 N) N q^2}{108 \mu ^2}+O\left(q^3\right).
\ee
For large $q$, we have
\bea
X&=&\frac{27}{4} \left(\frac{N-2}{N}\right)^{N-2}+\frac{27 \mu  (N-4) (N-2)^{N-3} N^{1-N}}{2 q}+O\left(\frac{1}{q^2}\right), \qquad N\ge 2\,,\nn\\
X&=& \ft{27}{16} \big(\fft{q}{\mu}\big)^{2-N}\Big((2-N)^{2-N} (3-N)^{N-3} N^2\nn\\
&&\qquad+\frac{N \left(N^3-3 N^2-4 N+8\right) \left(\frac{N-2}{N-3}\right)^{1-N} \mu}{(N-3)^2 q}
\Big)\,,\qquad 2-N\ge 0\,.
\eea
Note that when $N=2$ the two expansions become the same. Thus we have $X\ge 1$ for all the black hole parameters.  The case with $N=3$ is particularly simple, given by
\be
X= 1 + \frac{q (4 \mu +5 q)}{4 (\mu +q)^2}\,.
\ee
The analogous results can also be obtained for $Y$ and $Z$, which are both monotonic decreasing functions of $q$, for given $\mu$.  The minimum occurs at $q=0$, corresponding to $Y=1=Z$.  In particular, the $N=2$ and $N=3$ cases can be established manifestly, namely
\bea
N=2:&& Y=\fft12 + \frac{1}{2} \sqrt{1+\frac{8 q}{9 \mu }}\,,\qquad Z=1 + \frac{2 q \sqrt{9 \mu +8 q}}{9 \sqrt{\mu } \left(3 \mu +2q+\sqrt{\mu(9 \mu +8 q)}\right)}\,,\nn\\
N=3:&& Y=\sqrt{1 + \frac{2 q}{3 \mu +q}}\,,\qquad Z=\sqrt{1 + \fft{q}{3\mu}}\,.
\eea

\subsection{Multi-charged black holes}

In this subsection, we consider the multi-charge black holes in the STU supergravity model.  This black hole was also studied by \cite{Cvetic:2016bxi} in the context of (\ref{known1}) and (\ref{known2}). The asymptotically flat metric in the Einstein frame is given by
\cite{Behrndt:1996hu,Cvetic:1999xp}
\be
h=f=\fft{1 - \fft{\mu}{r}}{\sqrt{H_1 H_2 H_3 H_4}}\,,\qquad
R=r (H_1 H_2 H_3 H_4)^{\fft14}\,,\qquad H_i = 1 + \fft{q}{r}\,.
\ee
The solution contains five non-negative parameters $(\mu, q_1,q_2,q_3,q_4)$ parameterizing the mass and four types of $U(1)$ charges. The mass of the black hole is
\be
M=\ft12\mu + \ft14 (q_1 + q_2 + q_3 + q_4)\,.
\ee
The horizon is located at $r_+=\mu$ and the horizon radius is
\be
R_+=\prod_{i=1}^4 (\mu + q_i)^{\fft14}\,.
\ee
The location of the photon sphere $r_{\rm ph}$ satisfy the the quintic equation
\bea
&&r_{\rm ph}^4 \left(2 r_{\rm ph}-3 \mu \right) + \left(q_1+q_2+q_3+q_4\right) r_{\rm ph}^3 \left(r_{\rm ph}-2 \mu \right)\nn\\
&&-\mu  \left(q_1 q_2+q_3 q_2+q_4 q_2+q_1 q_3+q_1 q_4+q_3 q_4\right) r_{\rm ph}^2+q_1 q_2 q_3 q_4 \left(\mu -2 r_{\rm ph}\right)\nn\\
&&-\left(q_1 q_2 q_3+q_1 q_4 q_3+q_2 q_4 q_3+q_1 q_2 q_4\right) r_{\rm ph}^2=0\,.\label{condph}
\eea
The radii of the photon sphere and the shadow disk are
\be
R_{\rm ph} = \prod_{i=1}^4 (r_{\rm ph} + q_i)^{\fft14}\,,\qquad
R_{\rm sh} = \sqrt{\fft{\prod_{i=1}^4 (r_{\rm ph} + q_i)}{r_{\rm ph} (r_{\rm ph}-\mu)}}\,.
\ee
The proof of the inequalities become simpler when one of the charge parameter becomes zero. To illustrate this, we set, without loss of generality, $q_4=0$. We can solve for $\mu$ from (\ref{condph}) and substitute it into $(X,Y,Z)$ defined in (\ref{XYZ}).  For $X$, we find
\bea
X &=&1+  \fft{1}{16 r_{\rm ph}^4 (H_1 H_2 + H_1 H_3 + H_2 H_3)^3}\Big[
216 \left(q_1+q_2+q_3\right) r_{\rm ph}^5\nn\\
&&+9 \left(59 q_1^2+70 q_2 q_1+70 q_3 q_1+59 q_2^2+59 q_3^2+70 q_2 q_3\right) r_{\rm ph}^4\nn\\
&&+4 \big(103 q_1^3+219 q_2 q_1^2+219 q_3 q_1^2+219 q_2^2 q_1+219 q_3^2 q_1\nn\\
&&+240 q_2 q_3 q_1+103 q_2^3+103 q_3^3+219 q_2 q_3^2+219 q_2^2 q_3\big) r_{\rm ph}^3\nn\\
&&+6 \big(18 q_1^4+85 q_2 q_1^3+85 q_3 q_1^3+110 q_2^2 q_1^2+110 q_3^2 q_1^2+143 q_2 q_3 q_1^2+85 q_2^3 q_1+85 q_3^3 q_1\nn\\
&&\qquad +143 q_2 q_3^2 q_1+143 q_2^2 q_3 q_1+18 q_2^4+18 q_3^4+85 q_2 q_3^3 +110 q_2^2 q_3^2+85 q_2^3 q_3\big) r_{\rm ph}^2\nn\\
&&+12 \left(q_1+q_2+q_3\right) \big(9 q_2 q_1^3+9 q_3 q_1^3+10 q_2^2 q_1^2+10 q_3^2 q_1^2+11 q_2 q_3 q_1^2+9 q_2^3 q_1\nn\\
&&\qquad +9 q_3^3 q_1+11 q_2 q_3^2 q_1+11 q_2^2 q_3 q_1+9 q_2 q_3^3+10 q_2^2 q_3^2+9 q_2^3 q_3\big) r_{\rm ph}\nn\\
&& +27 q_1^2 q_2^4+27 q_1^2 q_3^4+27 q_2^2 q_3^4+54 q_1 q_2 q_3^4+38 q_1^3 q_2^3+38 q_1^3 q_3^3+38 q_2^3 q_3^3\nn\\
&&+60 q_1 q_2^2 q_3^3+60 q_1^2 q_2 q_3^3+27 q_1^4 q_2^2+27 q_1^4 q_3^2+27 q_2^4 q_3^2+60 q_1 q_2^3 q_3^2+93 q_1^2 q_2^2 q_3^2\nn\\
&&+60 q_1^3 q_2 q_3^2+54 q_1 q_2^4 q_3+60 q_1^2 q_2^3 q_3+60 q_1^3 q_2^2 q_3+54 q_1^4 q_2 q_3
\Big]\,.
\eea
Thus we see that the polynomial terms in the bracket all have positive coefficient and hence $X\ge 1$.  The same conclusion can be made for $Y$ and $Z$ as well. We thus demonstrate that the relations in (\ref{conjecture}) hold for all the three-charge configurations.

For the general $q_1 q_2 q_3 q_4\ne 0$, we do not have a clever analytic way to prove (\ref{conjecture}). We solved for $R_+, R_{\rm ph}, R_{\rm sh}$ and $M$ for given parameters $(\mu, q_i)$ in wide range and we found that (\ref{conjecture}) always holds.  Some special cases can be proven analytically. When $q_i$'s are equal, the solution reduces to the RN black hole for which (\ref{conjecture}) was proven in section \ref{sec:em}.
We find that when the charges are pair-wise equal, an analytical proof is also possible.  In the remainder of this section, we present the proof.  We set $q_3=q_1$ and $q_4=q_2$ and prove that $Z\ge 1$ first.  We define a function $W$, given by
\be
W(r)\equiv r^6 (H_1 H_2 H_3 H_4)^{\fft32} \Big(\fft{h}{R^2}\Big)'=-2 r^3+3 \mu  r^2
+r \left(\mu  q_2+q_1 \left(\mu +2 q_2\right)\right)-\mu  q_1 q_2\,.
\ee
The photon sphere is located at the largest root of $W$.  This implies that if there exists $W(r^*)>0$, then we must have $r_{\rm ph} > r^*$. We now consider a specific $r^*>0$, defined by
\be
r^*=\ft12 (v - q_1 - q_2)\,,\qquad v\equiv \sqrt{9 \mu ^2+9 \mu  \left(q_1+q_2\right)+q_1^2+q_2^2+7 q_1 q_2}\,.
\ee
We find that $W(r^*)=U+V$ where
\bea
U&=& \ft14 (\mu +q_1+q_2) \Big(27 \mu ^2+27 \mu  \left(q_1+q_2\right)+4 q_1^2+4 q_2^2+19 q_1 q_2\Big)\,,\nn\\
V&=& -\ft14 v \Big(9 \mu ^2+13 \mu  \left(q_1+q_2\right)+4 q_1^2+4 q_2^2+9 q_1 q_2\Big)\,.
\eea
Since $U$ is positive and $U^2-V^2$ is manifestly nonnegative, it follows that $W(r^*)\ge 0$, and hence $r_{\rm ph}\ge r^*$. We can now introduce a parameter $y\ge 0$ and write $r_{\rm ph}$ as
\be
r_{\rm ph}= \ft12 (\sqrt{v^2 + 9y(\mu +q_1)(\mu +q_2)} - q_1 - q_2)\ge r^*\,.
\ee
Substitute this into the expression for $Z$, we find that
\be
Z=1 + y\ge 1\,.
\ee
Having established that $Z\ge 1$, we can make use of this information to prove $(X,Y)\ge 1$. We define a dimensionless charge ratio $x=q_1/q_2$. The inequality $Z\ge 1$ implies that we can parameterize $r_{\rm ph}$ as
\be
r_{\rm ph}= \ft{1}{2} q_2\,\left(\sqrt{x^2+9 x y+7 x+1}-x-1\right), \qquad x\ge 0\,,\qquad y\ge 0\,.
\ee
We find
\bea
X&=& \frac{3 x (9 y+1)^2 (x (x+9 y+7)+1) \left(\sqrt{x (x+9 y+7)+1}-x-1\right)^2}{(y+1) \left(4 x^2+x (27 y+19)-4 (x+1) \sqrt{x (x+9 y+7)+1}+4\right)^3}\,,\nn\\
Y&=&\frac{4 x^2+x (27 y+19)-4 (x+1) \sqrt{x (x+9 y+7)+1}+4}{3 \left(\sqrt{x (x+9 y+7)+1}-x-1\right)^2}\,.
\eea
It is a simple exercise to show that $X\ge 1$ and $Y\ge 1$ for all nonnegative dimensionless parameters $(x,y)$.

\section{Einstein-Born-Infeld black hole}

The dyonic black hole of the Einstein-Born-Infeld (EBI) theory was constructed quite sometime ago in \cite{GSP1984}. The metric functions for the static solution are given by
\be
f=h=1-\frac{2 M}{r}+\frac{1}{6} b^2 \Big(r^2-\sqrt{\frac{Q^2}{b^2}+r^4}\Big) +\frac{Q^2 \, _2F_1\left(\frac{1}{4},\frac{1}{2};\frac{5}{4};-\frac{Q^2}{b^2 r^4}\right)}{3 r^2}\,,\qquad
R=r\,,
\ee
where $Q^2=Q_e^2 + Q_m^2$.  The metric  is asymptotic to the Minkowski spacetime in large $r$ and develops a curvature singularity at $r=0$.  The Taylor expansions of $f$ for large and small $r$ are
\bea
f&=& 1-\frac{2 M}{r}+\frac{Q^2}{4r^2}-\frac{Q^4}{80b^2 r^6}+{\cal O}\left(\frac{1}{r^{10}}\right);\nn\\
f &=& -\fft{2(M-M^*)}{r} + (1-\ft12 bQ) + \ft16 b^2 r^2 + {\cal O}(r^3),
\eea
where
\be
M^*=\frac{\Gamma \left(\frac{1}{4}\right) \Gamma \left(\frac{5}{4}\right)}{6 \sqrt{\pi }}
\sqrt{b Q^3}\sim 0.309 \sqrt{b Q^3}\,.
\ee
Black holes with only single horizon or two horizons can both emerge depending on the parameter of the solutions.  In order to demonstrate the validity of (\ref{conjecture}), we begin with a review of the discussion given in \cite{Li:2016nll}. When $bQ\le 2$, the solution describes a black hole only when $M>M^*$ and the black hole has only one horizon, analogous to the Schwarzschild black hole.  When $bQ> 2$, single-horizon black holes arise again for $M\ge M^*$. Furthermore two-horizon black holes can also emerge for $M_{\rm ext} \le M< M^*$, where
$M_{\rm ext}$ is the mass for the extremal solution, given by
\be
M_{\rm ext} = \frac{2 b^2 Q^2 \, _2F_1\left(\frac{1}{4},\frac{1}{2};\frac{5}{4};-\frac{16 b^2 Q^2}{\left(b^2 Q^2-4\right)^2}\right)+b^2 Q^2-4}{6 b \sqrt{b^2 Q^2-4}}\,.
\ee
The horizon radius of the extremal black hole is
\be
r_{\rm ext} = \frac{\sqrt{b^2 Q^2-4}}{2 b}\,.
\ee
Thus for $bQ> 2$, the solution describe a black hole if $M\ge M_{\rm ext}$ and the outer horizon radius satisfies $r_+\ge r_{\rm ext}$.   The locations of the outer horizon and photon sphere is determined by
\bea
f(r_+) &=&1-\frac{2 M}{r_+} -\frac{1}{6} b^2 \sqrt{\frac{Q^2}{b^2}+r_+^4}+\frac{1}{6} b^2 r_+^2+
\frac{Q^2 \, _2F_1\left(\frac{1}{4},\frac{1}{2};\frac{5}{4};-\frac{Q^2}{b^2 r_+^4}\right)}{3 r_+^2}=0\,,\nn\\
\Big(\fft{f}{r^2}\Big)'\Big|_{r_{\rm ph}} &=&\frac{6 M r_{\rm ph}-2 r_{\rm ph}^2- Q^2 \, _2F_1\left(\frac{1}{4},\frac{1}{2};\frac{5}{4};-\frac{Q^2}{b^2 r_{\rm ph}^4}\right)}{r_{\rm ph}^5}=0\,.
\label{ebicons}
\eea
We find that the quantities $(X,Y,Z)$ defined in (\ref{XYZ}) are given by
\bea
X &=& \frac{1}{8 r_{\rm ph}^4}\left(2+b^2 r_{\rm ph}^2-b \sqrt{b^2 r_{\rm ph}^4+Q^2}\right) \left(2 r_{\rm ph}^2+ Q^2 \, _2F_1\left(\frac{1}{4},\frac{1}{2};\frac{5}{4};-\frac{Q^2}{b^2 r_{\rm ph}^4}\right)\right)^2,\nn\\
Y&=& \frac{2}{2+b^2 r_{\rm ph}^2 -b \sqrt{b^2 r_{\rm ph}^4+Q^2}},\qquad Z=\fft{4r_{\rm ph}^2}{9 r_+^2}\,.
\eea
Note that for $bQ\le 2$, both of the two equations in (\ref{ebicons}) have only one real root for $r_+$ and $r_{\rm ph}$ respectively. On the other hand, when $bQ>2$, both equations can have two real roots and only the larger ones are relevant.

We now demonstrate the inequalities in (\ref{conjecture}).  The proof of $X\ge 1$, $Y\ge 1$ and $Z\ge 1$ are relatively simple when $bQ\le 2$. In this case, there is only one horizon and furthermore, the horizon radius $r_+$ and also $r_{\rm ph}$ can approach zero. It can be easily established that both $X$ and $Y$ are monotonically decreasing function of $r_{\rm ph}$, with
\bea
X&=& (2-b Q)\Big(\frac{2 b Q^3 \Gamma(\fft{5}{4})^4}{\pi  r_{\rm ph}^2}-\frac{\sqrt{b Q^3} \Gamma (\fft{5}{4})^2 (b Q-2)}{\sqrt{\pi } r_{\rm ph}}+{\cal O}\left( r_{\rm ph}^0\right)
\Big),\nn\\
X&=&1+\frac{3 Q^2}{4 r_{\rm ph}^2}+{\cal O}\left(\frac{1}{r_{\rm ph}^6}\right),\nn\\
Y&=& \frac{2}{2-b Q}-\frac{2 b^2 r_{\rm ph}^2}{(2-b Q)^2}+{\cal O}\left(r_{\rm ph}^3\right),\qquad
Y= 1+\frac{Q^2}{4 r_{\rm ph}^2}+\frac{Q^4}{16 r_{\rm ph}^4}+{\cal O}\left(\frac{1}{r_{\rm ph}^6}\right),
\eea
It thus becomes straightforward to see that $X\ge 1$ and $Y\ge 1$ for $bQ\le 2$.

The demonstration of $Z\ge 1$ is more subtle. We can solve for $M$ from the second equation in (\ref{ebicons}) and substitute it into $f(r)$.  We find that
\bea
f(\ft32 r_{\rm ph}) &=& \fft{Q^2}{4r_{\rm ph}^2}\left(3 \, _2F_1\left(\frac{1}{4},\frac{1}{2};\frac{5}{4};-\frac{81 Q^2}{16 b^2 r_{\rm ph}^4}\right)-2 \, _2F_1\left(\frac{1}{4},\frac{1}{2};\frac{5}{4};-\frac{Q^2}{b^2 r_{\rm ph}^4}\right)\right)\nn\\
&&+\ft{1}{54} b \left(4 b r_{\rm ph}^2-\sqrt{16 b^2 r_{\rm ph}^4+81 Q^2}\right).\label{ebif32rph}
\eea
It is easy to verify that the above quantity is positive for all positive $r_{\rm ph}$.  In particular, we have
\bea
f(\ft32 r_{\rm ph}) &=& \frac{2}{27} b^2 r_{\rm ph}^2-\frac{97 b^3 r_{\rm ph}^4}{1620 Q}+O\left(r_{\rm ph}^5\right)\,,\nn\\
f(\ft32 r_{\rm ph}) &=&\frac{Q^2}{16 r_{\rm ph}^2}-\frac{473 Q^4}{5120 b^2 r_{\rm ph}^6}+O\left(\frac{1}{r_{\rm ph}^{10}}\right).
\eea
Thus when the solution has only one horizon, the horizon radius $r_+$ must be less than $\fft32 r_{\rm ph}$, since $f(r_+)=0$ separates the $f(r)<0$ and $f(r)>0$ regions. As we can see, the quantity $f(\fft32 r_{\rm ph})$ is positive regardless the value of $bQ$, but as we shall see in a momentum that this in itself does not imply that $Z\ge 1$ must be true also for black holes with two horizons, since $f(r)$ inside the inner horizon is also positive.

When $bQ> 2$, the above discussion breaks down and it appears as if the inequalities are no longer valid.  That is in fact not true. As was discussed earlier, black hole with two horizons can emerge and there is a minimum for $r_+$ and hence $r_{\rm ph}$ for given $bQ>2$.  However, the analysis becomes very complicated since now both equations in (\ref{ebicons}) can have two real roots and we have to pick the bigger roots, but neither equations have analytical solutions. Although the expression (\ref{ebif32rph}) is always positive, independent of $bQ$, this does not necessarily guarantee $Z\ge 1$ when there are two horizons since now two possibilities could arise.  One is that $\fft32 r_{\rm ph}>r_+$ and the other is that it could be less than the inner horizon $r_-\le r_+$, where $f(r)$ is positive also.

In order to prove (\ref{conjecture}) for $bQ>2$, we have to make use of the fact that $r_+\ge r_{\rm ext}$.
We first establish $Z\ge 1$.  We can solve for $M$ from $f(r_+)=0$ and substitute into
\be
W(r) = \Big(\fft{f}{r^2}\Big)' = -\fft{2}{r^3} + {\cal O}\Big(\fft{1}{r^4}\Big).
\ee
Since the photon sphere corresponds to the largest root of $W$, it follows that if $W(r)>0$, then it is necessarily the case that $r<r_{\rm ph}$.  We define a dimensionless quantity $x$ by $r_+ = x r_{\rm ext}$ such that $x\ge 1$. We evaluate
\be
\widetilde W(x) = W(\ft32 r_+)=W(\ft32 x r_{\rm ext})\,.
\ee
We find that $\widetilde W(x)$ is a monotonically decreasing function from $x=1$ to $x=\infty$, with
\bea
\widetilde W(1) &=& -\frac{128 b^3}{81 \left(b^2 Q^2-4\right)^{3/2}} +\frac{512 b^5 Q^2}{243 \left(b^2 Q^2-4\right)^{5/2}}\nn\\
&&\times \Bigg( 3 \, _2F_1\left(\frac{1}{4},\frac{1}{2};\frac{5}{4};-\frac{16 b^2 Q^2}{\left(b^2 Q^2-4\right)^2}\right)-2 \, _2F_1\left(\frac{1}{4},\frac{1}{2};\frac{5}{4};-\frac{256 b^2 Q^2}{81 \left(b^2 Q^2-4\right)^2}\right)\Bigg)\nn\\
\widetilde W(x) &=& \frac{128 b^5 Q^2}{243 x^5 \left(b^2 Q^2-4\right)^{5/2}} + {\cal O}\Big(\fft{1}{x^9}\Big)\,.
\eea
The quantity $\widetilde W(1)$ is itself a positive monotonic decreasing function of $b Q$, running from $bQ=2+$ to $bQ\rightarrow \infty$. We thus demonstrate $W(\ft32 r_+)\ge 0$, which implies that $r_{\rm ph} \ge \fft32 r_+$. We therefore have $Z\ge 1$.

    We can now make use of the proved fact $Z\ge 1$ and define $r_{\rm ph} = \fft32 y r_{\rm ext}$, then we must have $y\ge 1$.  This allows us to prove that $X\ge 1$ and $Y\ge 1$, since both are monotonically decreasing functions of $y$, namely
\bea
&&1+\frac{4 b^2 Q^2}{3 y^2 \left(b^2 Q^2-4\right)}+{\cal O}\left(\frac{1}{y^4}\right)\le
X\le X|_{y=1}\,,\nn\\
&&1+\frac{4 b^2 Q^2}{9 y^2 \left(b^2 Q^2-4\right)}+{\cal O}\left(\frac{1}{y^4}\right)\le Y \le Y|_{y=1}\,,
\eea
where
\bea
X|_{y=1} &=& \frac{1}{2592 \left(b^2 Q^2-4\right)^2}
\left(9 b^2 Q^2-4-\sqrt{\left(4-9 b^2 Q^2\right)^2-320 \left(b^2 Q^2\right)}\right)\nn\\
 &&\times \left(8 b^2 Q^2 \, _2F_1\left(\frac{1}{4},\frac{1}{2};\frac{5}{4};-\frac{256 b^2 Q^2}{81 \left(b^2 Q^2-4\right)^2}\right)+9 b^2 Q^2-36\right)^2\nn\\
Y|_{y=1} &=& \frac{32}{9 b^2 Q^2-4-\sqrt{\left(9 b^2 Q^2-4\right)^2-320 \left(b^2 Q^2-4\right)}}\,.
\eea
We therefore prove that the inequality relations in (\ref{conjecture}) are all satisfied by the dyonic black holes in the EBI theory.

\section{Further examples}

\subsection{Bardeen black hole}

We include the Bardeen's black hole \cite{bardeen} since it satisfies the weak energy condition, but not the dominant energy condition. Furthermore the trace of the energy-momentum tensor is positive.  These violate the conditions for the theorems (\ref{known1}) and (\ref{known2}), but we shall see that the sequences of inequalities in (\ref{conjecture}) all hold nevertheless. The metric functions of the Bardeen black hole are given by
\be
h=f=1 - \fft{2M r^2}{(r^2 + q^2)^{\fft32}}\,,\qquad R=r\,.
\ee
It should be pointed out that this is not the most general solution for the given matter energy-momentum tensor, since we can always add a term $\mu/r$ associated with the condensation of massless graviton in the above and the metric will satisfy the exact same equations of motion.  For sufficiently large $M$, the metric describes a black hole with no curvature singularity. Typically the black hole have two horizons, which coalesce in the extremal limit.  The mass and the horizon radius of the extremal solution is
\be
M_{\rm ext}= \ft34\sqrt3\,q\,,\qquad r_{\rm ext}=\sqrt2\,q\,.
\ee
For given $q$, the general black hole has $M\ge M_{\rm ext}$ with the outer horizon $r_+\ge r_{\rm ext}$.
The black hole thermodynamics was studied recently in \cite{Zhang:2016ilt}.

The radius of the photon sphere is determined by
\be
\frac{6 M r_{\rm ph}}{(r_{\rm ph}^2+q^2)^{\fft52}}-\frac{2}{r_{\rm ph}^3}=0\,.
\ee
Thus the quantities $(X,Y,Z)$ defined in (\ref{XYZ}) are given by
\be
X=\frac{\left(r_{\rm ph}^2-2 q^2\right) \left(r_{\rm ph}^2+q^2\right)^5}{r_{\rm p}^{12}},\qquad Y=\frac{r_{\rm ph}^2}{r_{\rm ph}^2-2 q^2},
\qquad Z=\frac{4 r_{\rm ph}^2}{9 r_+^2}\,.
\ee
We now follow the same technique developed earlier and prove that $Z\ge 1$ first.  In fact, this simpler but nontrivial example may provide an easier illustration of the technique. We can solve for $M$ from $f(r_+)=0$ and then obtain
\be
W(r)\equiv \Big(\fft{f}{r^2}\Big)'=\frac{3 r \left(q^2+r_+^2\right)^{3/2}}{r_+^2 \left(q^2+r^2\right)^{5/2}}-\frac{2}{r^3}\,.
\ee
We define the dimensionless parameter $x\ge 1$ by $r_+=x r_{\rm ext}$. We find that for $x\ge 1$, we have
\be
W(\ft32 r_+) =\frac{\sqrt{2}}{27 q^3 x^3}\Big( \frac{243 \sqrt{2} x^2 \left(2 x^2+1\right)^{3/2}}{\left(9 x^2+2\right)^{5/2}}-4\Big)\ge 0\,.
\ee
Since $r_{\rm ph}$ is the largest root of $W$, it follows that $r_{\rm ph}\ge\ft32 r_+$, implying $Z\ge 1$.  This also implies that if we define $r_{\rm ph} = \fft32(1+y) r_{\rm ext}$, then we must have $y\ge 0$.  Substituting this redefinition of $r_{\rm ph}$ to $(X,Y)$, we find
\bea
X-1 &=&\frac{2}{531441 (y+1)^{12}} \Big(177147 y^{10}+1771470 y^9+7971615 y^8\nn\\
&&+21257640 y^7 +37171710 y^6+44466084 y^5+36753750 y^4\nn\\
&&+20635560 y^3 +7474599 y^2+1555038 y+136907\Big)\ge 0,\nn\\
Y-1 &=&\frac{4}{(3 y+1) (3 y+5)}\ge 0\,.
\eea
Therefore the sequence of the inequalities in (\ref{conjecture}) are all satisfied by this black hole.

\subsection{Black hole in quasi-topological electromagnetism}

In all the previous examples, black holes have only single or two horizons and furthermore there is only one (unstable) photon sphere outside the outer horizon. Recently explicit black holes with four horizons were also constructed, satisfying the dominant energy condition, in the Einstein-Maxwell theory extended with quasi-topological electromagnetism \cite{Liu:2019rib}.  For appropriate mass, the four horizons can reduce to two and the metric function $h=f$ can have a wiggle outside the outer horizon such that the black hole can have three photon spheres, with a stable one sandwiched between two unstable ones.

The metric function of the black hole is expressed in terms a hypergeometric function, given by \cite{Liu:2019rib}
\be
h=f=1-\frac{2 M}{r}+\frac{\alpha _1 p^2}{r^2}+\frac{q^2 \, _2F_1\left(\frac{1}{4},1;\frac{5}{4};-\frac{4 p^2 \alpha _2}{r^4 \alpha _1}\right)}{\alpha _1 r^2}\,,\qquad R=r\,.
\ee
Here $(\alpha_1,\alpha_2)$ are respectively the coupling constants of kinetic and quasi-topological terms of the Maxwell field.  In this paper, we shall not attempt to prove the inequality relations in (\ref{conjecture}) for this case, since it is exceedingly complicated when we have to deal with four horizons with the metric function involving a hypergeometric function.  Instead we shall look at a specific example of three photon spheres and their shadows.

The mass, electric and magnetic charges are given by $M$, $q$ and $p$ respectively.
Choosing a specific set of  parameters
\be
(M, q^2,p^2,\alpha_1, \alpha_2)=(\ft{67}{10}, \ft{20868}{443},\ft{396}{443},1,\ft{196249}{1584})\,,
\ee
The metric function becomes \cite{Liu:2019rib}
\be
f=1-\fft{67}{5r} + \fft{12}{443r^2} \big(33 + 1739 \, _2F_1[\ft{1}{4},1;\ft{5}{4};-\ft{443}{r^4}]\big)\,.
\ee
One can use numerical plot to determine that the inner and outer horizon are located at
\be
r_-=0.6682\,,\qquad r_+=1.5498\,,\,.
\ee
The three photon spheres are located at \cite{Liu:2019rib}
\be
r_{\rm ph}=\{12.4354, 6.4430, 2.1939\}
\ee
What's interesting is that for this black hole, the photons will be trapped in the thick spherical shell between radius 2.1939 and 12.4354 and they will conglomerate at the stable photon sphere of radius 6.4430.  In other words, the photons captured by the black hole may not fall into the event horizon, but form a stable black hole photon shield. This phenomenon will not be observed from the asymptotic infinity, where one only sees a shadowy disk of radius
\be
r_{\rm sh}=25.8335.
\ee
We see that the relations in (\ref{conjecture}) all hold for this specific black hole also.

\section{Conclusions}

In this paper, we considered spherically symmetric and static black holes in Einstein gravity and compared the various parameters that characterize the black hole size. These are the radii of the event horizon, the photon sphere and the shadowy disk observed from asymptotic infinity. We found that they satisfied a sequence of inequalities summarized in (\ref{conjecture}).  We tested successfully the validity of (\ref{conjecture}) using a diverse selection of black holes, satisfying at least the null energy condition, suggesting that there should be some underlying principle for this set of inequality relations. These include the inequalities (\ref{known1}) and (\ref{known2}) which were proven in literature for black holes satisfying dominant energy condition, together the negative trace of the energy-momentum tensor. However they remain conjectures for black holes satisfying the null energy condition.

If the radius of the horizon measures the {\it actual} size of a black hole, photon spheres and shadowy disks give the {\it apparent} sizes. The Schwarzschild black hole saturates all the inequalities; therefore, for given mass, by both contents and appearances, it is the biggest of all.

There are also applications of our inequality relations in astrophysical observations. Our inequality $\fft32 R_+\le R_{\rm ph}$ indicates that the photon sphere has a bigger magnification than the naive expectation $R_+\le R_{\rm ph}$. The shadows we observe may not necessarily be that of a black hole but instead a massive object of the same mass with the size more than $R_+$ but less than $\fft32 R_{+}$.  Furthermore, our inequality $\fft32 R_+\le R_{\rm ph}$ tightens the bound $\Omega_{\infty} R_+\le 1/\sqrt3$ proposed in \cite{Hod:2017xkz} to $\Omega_{\infty} R_+\le 2/(3\sqrt3)$.  Finally, both mass $M$ and shadow size $R_{\rm sh}$ are measurable quantities through astrophysical observation, the upper bound $R_{\rm sh}\le 3\sqrt3 M$ provides a good verifiable test of Einstein's theory of gravity.

We also reexamined the black hole with three photon spheres, a stable one sandwiched between two unstable ones, first constructed in \cite{Liu:2019rib}.  The black hole satisfies the dominant energy condition and hence the existence of such black holes cannot be ruled out. The unusual feature is that the black hole can capture photons in a thick spherical shell outside the event horizon. We would like to call this stable surrounding aura as photon shield.  The outer unstable photon sphere will necessarily cast a shadow and the relevant parameters also satisfy (\ref{conjecture}). The shadow hides the photon shield from an observer at infinity.

There are two directions we can generalize our proposal (\ref{conjecture}).  One is to consider spherically symmetric and static black holes in general higher dimensions and we expect that relations become
\be
\big(\ft12(D-1)\big)^{\fft{1}{D-3}}R_+\,\le R_{\rm ph}\,\le
\sqrt{\ft{D-3}{D-1}}\, R_{\rm sh}\,\le\,
\Big(\ft{8\pi M(D-1)}{(D-2)\Omega_{D-2}}\Big)^{\fft{1}{D-3}}\,.
\ee
where $\Omega_{D-2}$ is the volume of the unit round $S^{D-2}$ and $M$ is the mass, entering the metric function as
\be
h=1 - \fft{16\pi M}{(D-2)\Omega_{D-2}}\,\fft{1}{R^{D-3}} + \cdots\,.
\ee
Indeed we verified the inequalities using the RN black holes in general $D$ dimensions, following the same technique of section \ref{sec:em}. A more general proof is still lacking.  The $R_+$ and $M$ relation implies an upper bound for the black hole entropy, namely
\be
S\le \fft14 \Omega_{D-2}\,\Big(\fft{16\pi M}{(D-2)\Omega_{D-2}}\Big)^{\fft{D-2}{D-3}}.
\ee
This presumably is the upper bound of entropy for any quantum system of energy $E=M$.

The other direction is to generalize the relations to include rotating black holes for which the shapes of both the photon surfaces and shadows become less regular.  Regardless of the outcome, the robustness of our proposal (\ref{conjecture}) for spherically-symmetric black holes satisfying null energy conditions calls for an abstract proof to better understand the underlying principle. If a counterexample could be indeed found, it is still of interest to understand its distinguishing feature from those that do obey the inequalities.

\section*{Acknowledgement}

This work is supported in part by NSFC Grants No.~11875200 and No.~11935009.

\end{document}